\documentclass{article}
\usepackage{spconf,amsmath,graphicx}

\usepackage{cite}
\usepackage{amssymb,amsfonts}
\usepackage{algorithmic}
\usepackage{graphicx}
\usepackage{textcomp}
\usepackage{xcolor}

\usepackage{algorithmic}
\usepackage{multirow}
\usepackage{threeparttable}
\usepackage{booktabs}
\usepackage{tabularx} 
\usepackage{colortbl}  
\usepackage{bm}
\usepackage{hyperref}
\usepackage{url}

\title{FINT: Field-aware INTeraction Neural Network For CTR Prediction}
\name{Zhishan Zhao$^{1}$, 
Sen Yang $^{2,3}$, 
Guohui Liu$^{1}$, 
Dawei Feng$^{2,3}$, 
Kele Xu$^{2,3}$\sthanks{Corresponding author.}
}
\address{      
	$^1$ iQIYI Inc., China\\ 	
	$^2$ National Key Lab of Parallel and Distributed Processing, Changsha, China\\
	$^3$ National University of Defense Technology, Changsha, China\\
	zhishan777@gmail.com, kelele.xu@gmail.com
}

\begin{document}
%
\maketitle
\begin{abstract}
As a critical component for online advertising and marketing, click-through rate (CTR) prediction has drawn lots of attention from both industry and academia. Recently, deep learning has become the mainstream methodological choice for CTR. Despite sustainable efforts have been made, existing approaches still pose several challenges. On the one hand, high-order interaction between the features is under-explored. On the other hand, high-order interactions may neglect the semantic information from the low-order fields. In this paper, we proposed a novel prediction method, named FINT, that employs the Field-aware INTeraction layer which explicitly captures high-order feature interactions while retaining the low-order field information. To empirically investigate the effectiveness and robustness of the FINT, we perform extensive experiments on the three realistic databases: KDD2012, Criteo and Avazu. The obtained results demonstrate that the FINT can significantly improve the performance compared to the existing methods, without increasing the amount of computation required. Moreover, the proposed method brought about 2.72\% increase to the advertising revenue of iQIYI, a big online video app through A/B testing.
To better promote the research in CTR field, we released our code as well as reference implementation at: https://github.com/zhishan01/FINT.
\end{abstract}
\begin{keywords}
Click-through Rate, Field-Aware Interaction, High-order Feature Interactions.
\end{keywords}
\section{Introduction}
	Click-through rate (CTR) prediction aims to forecast the probability that a user will click a particular recommended item or an advertisement on a web page \cite{richardson2007predicting}. The applications of CTR seem to be evident in several different fields, such as recommendation systems, online advertising and product search.
    During the last decades, CTR has drawn dramatic interest, due to its important roles in both the academic and industry. Unlike other data types, such as images and texts, data used in CTR are usually of high sparsity and large scale. Making an accurate and robust prediction is still far from being solved. In the early years, logistic regression (LR) and factorization machines (FM) were widely explored.
	Recently, deep learning-based approaches have been the mainstream methodological choice for CTR, such as, Wide\&Deep \cite{cheng2016wide}, DeepFM \cite{guo2017deepfm}, DCN \cite{wang2017deep} and xDeepFM \cite{lian2018xdeepfm}. Based on previous studies, how to fully utilize both the low- and high-order feature interactions simultaneously can bring extra performance improvements, compared to the cases of considering either alone.
	
	Existing high-order feature interaction based approaches still confront with a significant challenge. The field-level semantic information is lost during the high-order feature interaction. Consequently, the subsequent deep modules cannot fully employ the explicit features. In this paper, we propose a novel framework to capture high-order feature interactions while preserving the low-order field semantic information by introducing a field-aware interaction layer in the model. Specifically, a unique feature interaction vector is maintained for each field, which contains interaction information of the field and each other fields in any order within K and allow the subsequent DNN model to better explore the non-linear high-order relationship between fields. We employed this method to improve the CTR prediction accuracy on an online recommendation service, a system that recommends ads to users at iQIYI, which is one of the most popular online video apps in China. Extensive experiments on both public datasets and online A/B testing demonstrate the effectiveness and robustness of the proposed method.
		
	The remainder of this paper is structured as follows. In Section 2, we elaborately describe the proposed methodology. In Section 3, we perform comprehensive experiments on three widely used databases. Finally, in Section 4, we draw conclusions.
	
	\begin{figure*}[!htbp]
		\centering
		\includegraphics[width=0.9\textwidth]{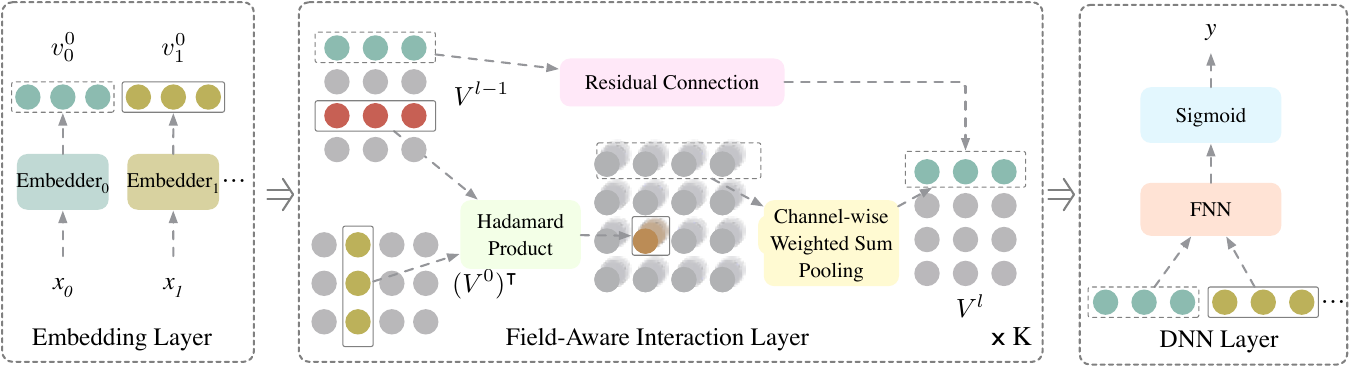}
		\caption{The architecture of FINT, which consists of: embedding layer, Field-Aware Interaction Layer, and DNN layer.}
		\label{fig.fint}
	\end{figure*}

	\section{Approach}
	We take CTR as a binary classification task to predict whether a user will click a given item. Specifically, we denote the model input as $X=\{x_0,\cdots,x_{M-1}\}$ which contains $M$ features. $X$ includes not only user features but also item features and context features. Such features could be categorical, such as age and gender, or continuous, such as item price. The gold target of CTR model is a scalar $y$ ($y=1$ means the user will click the given item, otherwise $y=0$). 
	
	\subsection{FINT}
	Figure~\ref{fig.fint} shows the overall architecture of our proposed model FINT. It mainly contains three modules: embedding layer, field-aware interaction layer, and DNN layer (dense neural network). The first one aims to embed features into dense vectors. The second one maintains a unique interaction vector for each field, it captures high-order field-aware feature interactions while retaining the low-order field information. The last module targets to exploit non-linear high order feature interaction and predict the user's clicking behavior.
	
	\textbf{Embedding Layer:} The FINT first embeds each feature $x_i\in X$ into a dense vector $v_i^0$ of dimension $D$ as its initial representation. If the field is multivalent, the sum of feature embedding is used as the field embedding. Here, "0" in the subscript denotes the initial one. 
 We denote the output of embedding layer, i.e., the initial representation of $X$, as:
\begin{equation}
V_0 = [v^0_{0},v^0_{1},\cdots,v^0_{M-1}]^\intercal.
\end{equation}

\textbf{Field-aware Interaction Layer:}
This layer aims to promote field-aware interaction between features to explore more possible combinations. Field-awareness means maintaining a unique interaction vector for each field, making further non-linear interaction among fields possible. We stack $K$ field-aware interaction layers in the FINT to achieve high-order feature interaction.
As shown in Figure~\ref{fig.fint}, to achieve field-aware interaction, each interaction layer conducts two steps, computing Hadamard product and channel-wise weighted sum pooling. The first step takes the initial representation $V^{0}$ and representation of the last step $V^{l-1}$ as input and computes the Hadamard product between all pairs of $v^{l-1}_i\in V^{l-1}$ and $v^0_j\in V^{0}$.
Then channel-wise weighted sum pooling is conducted along features with the learnable parameter $W^l\in\mathbb{R}^{M\times M}$.
To avoid training collapse, we also add residual connection~\cite{resnet} in each field-aware interaction layer. Specifically, representation $v_i^l$ is computed as:
	\begin{equation}
		v_{i}^{l} = \sum_{j=0}^{M-1} w_{i,j}^l\times(v_i^{l-1}\odot v_j^{0})+u^{l}_{i}v_i^{l-1},
		\label{equ.fil}
	\end{equation}
	where, $w^l_{i,j}$ is the element of $W^l$, $u_i^l$ is a scalar, $\odot$ denotes Hadamard product. Equation~\ref{equ.fil} indicates $v_i^l$ contains combination information of field $i$ and other fields in any order within $l$. If we consider all representations, Equation~\ref{equ.fil} can be re-organized in the matrix format:
	\begin{equation}
		V^l=V^{l-1}\odot (W^l\otimes V^0) + U^l\times V^{l-1},
		\label{equ.mfil}
	\end{equation} 
	in which $U^l=[u_0^l,\cdots,u_{M-1}^l]^\intercal$ is a learnable vector parameter for residual connection, $\otimes$ indicates matrix multiplication.
	
	\textbf{DNN Layer:} As the last module of FINT, the DNN layer is to explore non-linear interactions and predict the clicking probability according to the final representations. Concretely, the DNN layer works as:
	\begin{equation}
		\hat{y}=(\sigma \circ \text{FFN}) (v^K_0 \| \cdots \| v^K_{M-1}),
	\end{equation}
	where $\|$ indicates vector concatenation, $\circ$ indicates function composition, $\sigma$ is the sigmoid function, $\text{FFN}$ is a feed-forward network that contains multiple fully-connection layers with hidden size $D_F$ and active function RELU. Such an architecture allows the DNN layer to explore high-order non-linear feature interaction in the semantic space. In the training stage, we use binary cross entropy as the training loss for FINT. In the inference stage, we take the output $\hat{y}$ as the probability of a user clicking the given item.
	
	\textbf{Time Complexity:}
	Equation~\ref{equ.mfil} shows that each feature interaction layer can be efficiently computed in $\mathcal{O}(M^2 D)$. For the DNN layer, the vector-matrix multiplication is the main operation which can be done in $\mathcal{O}(MDD_F+D_F^2)$. Since there are $K$ field-aware interaction layers($K$ is usually small), the overall time complexity of FINT is $\mathcal{O}(KM^2 D+MDD_F+D_F^2)$. Although the feature interaction layer complexity is inferior to the one of traditional machine learning based FM~\cite{rendle2010factorization} and NFM~\cite{he2016deep}, which is $\mathcal{O}(KMD)$, it surpasses a variety of deep learning based peers, such as xDeepFM~\cite{guo2017deepfm,lian2018xdeepfm} which is $\mathcal{O}(KM^2DT)$, where $T$ is the number of multiple pooling operations. Moreover, as FINT conducts most operations on matrixes and requires no sequence operations, it can achieve a higher GPU acceleration ratio.
	
	\subsection{Relationship with Other Models}
	FINT shares a similar paradigm with several factorization based models~\cite{rendle2010factorization,he2016deep,lian2018xdeepfm,song2019autoint}, as they explore feature interaction in the vector space and exploit pooling operations to reduce dimension and promote further classification. 
	NFM~\cite{he2017neural} has shown the advantage of integrating linear feature combination and nonlinear high-order feature combination. 
	Therefore, FINT also exploits nonlinear feature interaction through the DNN layer.
	On the other hand, the field-aware interaction layer makes FINT distinguishable from others. 
	Unlike with previous approaches that cast all feature representations into a single scalar or vector during feature interaction~\cite{rendle2010factorization,he2016deep,lian2018xdeepfm}, the field-aware interaction layer maintains a vector for each field, retaining their boundaries, to allows the following DNN module to further mine nonlinear interaction.
	AutoINT~\cite{song2019autoint} is the most related work to FINT in paradigm because it also retains the feature boundary in linear feature interaction and exploits nonlinear high-order feature interaction. However, it is based on the Transformer model~\cite{vaswani2017attention} and uses the self-attention mechanism to learn feature weights, while FINT uses the Hadamard product, which is a more general and effective method in recommendation systems.

	\section{EXPERIMENTS}
	In this section, we aim to provide the experimental results of FINT, from the perspective of efficiency and effectiveness. The prototype of FINT is implemented by TensorFlow 1.14.0 and runs with a Nvidia Tesla P40 GPU. Two metrics Logloss and AUC (Area Under the ROC Curve) are used in our experimental studies. For the training of the FINT model, we set the learning rate as 1e-3 and Adam optimizer is employed in our experiments. The batch size is set 1024, while the embedding size is set as 16. We use 3 field-aware interaction layers. For the DNN layers, the number of hidden layers is set as [300, 300, 300]. We replace the features that appear less than 10 times as ``unknown''. Numerical features can be normalized with $z^* = log(z+1) + 1$. The settings of the experimental part are basically kept the same as AutoINT.

	\subsection{Dataset and Baselines}
	We evaluate the proposed method on three publicly available datasets including the KDD12 \footnote{https://www.kaggle.com/c/kddcup2012-track2},  Criteo \footnote{https://www.kaggle.com/c/criteo-display-ad-challenge}, and Avazu \footnote{https://www.kaggle.com/c/avazu-ctr-prediction}. Criteo and Avazu contain chronologically ordered click-through records from Criteo and Avazu which are two online advertising companies. For the Avazu and Cretio dataset, we randomly split the dataset into training (80\%), validation (10\%), and test (10\%) sets. While for the KDD12, we follow the official public and private split. 
	We implemented 9 widely used CTR prediction approaches, and compared them in the experiment. Below is a brief introduction of these models:

\textbf{LR}, we employ the LR only with basic features as our first baseline.
 
\textbf{FM}, we employ the original FM model, which has demonstrated its effectiveness in many CTR prediction tasks.

\textbf{NFM}, which aims to encode all feature interactions, through a multi-layer neural network coupled with a bit-wise bi-interaction pooling layer.

 \textbf{PNN}, which applies a product layer and multiple fully connected layers to explore the high-order feature interactions.
 
 \textbf{Wide \& Deep}, which aims to model low- and high-order feature interactions simultaneously.
 
\textbf{DeepFM}, which explores the integration of the of FM and deep neural networks (DNN). Through the modeling of low-order feature interactions like FM and models high-order feature interactions like DNN.

 \textbf{AutoInt}, which employs a multi-head self-attentive neural network as the core module and can automatically learn the high-order interactions of input features.
 
 \textbf{DCN}, which makes use of the deep cross network and takes the outer product of concatenated feature embeddings to explicitly model feature interaction.
 
 \textbf{xDeepFM}, which has a compressed interaction network to model vector-wise feature interactions for CTR prediction.
		\begin{table*}[]
		\centering
		\caption{Effectiveness comparison of different algorithms.}
		\begin{tabular}{m{1.8cm}<{\centering}m{1.6cm}<{\centering}m{1.6cm}<{\centering}m{1.6cm}<{\centering}m{1.6cm}<{\centering}m{1.6cm}<{\centering}m{1.6cm}<{\centering}}
			\hline
			\multirow{2}{*}{Model} &
			\multicolumn{2}{c}{Criteo} &
			\multicolumn{2}{c}{Avazu} &  \multicolumn{2}{c}{KDD12}\\
			&AUC&Logloss&AUC&Logloss &AUC&Logloss \\
			\hline
			LR & 0.7846 & 0.4670 & 0.7616 & 0.3901 & 0.7352& 0.1385 \\
			FM & 0.7912 & 0.4627 & 0.7753 & 0.3826 & 0.7419& 0.1383 \\
			NFM & 0.7991 & 0.4541 & 0.7761 & 0.3820 & 0.7419 & 0.1378 \\
			PNN & 0.8069 & 0.4473 & 0.7793 & 0.3802 & 0.7571 & 0.1357 \\
			DeepFM & 0.8014 & 0.4524 & 0.7785 & 0.3806 & 0.7517 & 0.1366 \\
			Wide/ Deep & 0.8042 & 0.4495 & 0.7776 & 0.3811&0.7509 &  0.1366\\
			AutoInt & 0.8053 & 0.4482 & 0.7770 & 0.3813 & 0.7613&0.1356 \\
			DCN & 0.8053 & 0.4483 & 0.7777 & 0.3811 & 0.7546 & 0.1360\\
			XDeepFM & 0.8055 & 0.4484 & \textbf{0.7796} & 0.3801 & 0.7531 & 0.1360 \\
			\hline
			FINT & \textbf{0.8077} & \textbf{0.4461} & 0.7795 & \textbf{0.3800} & \textbf{0.7618} & \textbf{0.1355} \\
			\hline
		\end{tabular}
		\label{tab.result}
	\end{table*}

\subsection{Performance Evaluation}
	As can be observed from Table~\ref{tab.result}: the proposed FINT model can provide better performance compared to other models on Criteo, KDD2012, and Avazu on both metrics. The FINT obtains different improvements on three datasets. For example, on the Criteo dataset, FINT surpasses the previous best model (PNN) over 0.08\% point on AUC, which is already a considerable improvement in the CTR task.  On the Avazu dataset, the FINT achieves a comparable AUC with XDeepFM, only superior with 0.01\% point. On the other hand, it outperforms XDeepFM on the Logloss metric. The improvement on the KDD2012 dataset is similar to the one of Criteo.

	\begin{figure}[!htbp]
		\centering
		\includegraphics[scale=0.5]{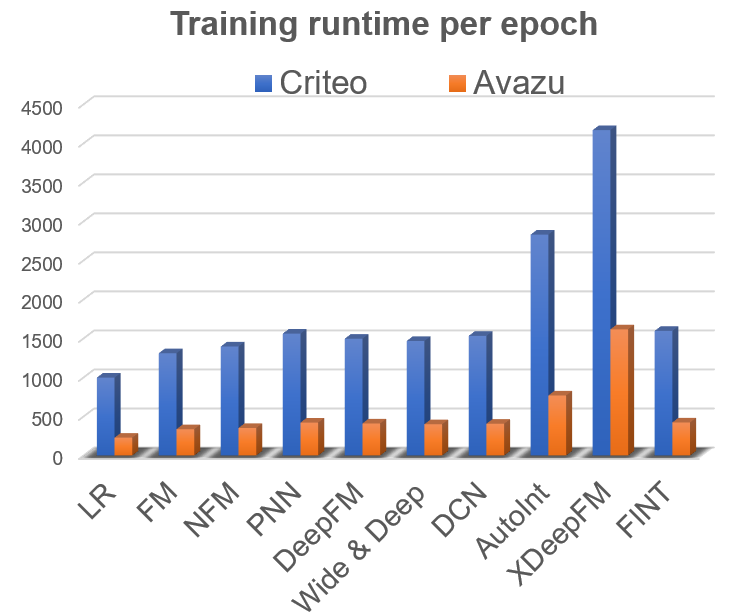}
		\caption{Time comparison (in seconds).}
		\label{fig.eff}
	\end{figure}

	\subsection{Effectiveness Comparison}
	In Figure~\ref{fig.eff}, we conducted a quantitative comparison on the runtime
	between FINT and seven state-of-the-art models with GPU implementations on Criteo and Avazu. In the Figure, the y-axis provides an average runtime per epoch over five training epochs after which all models start to converge observably. We keep the hardware settings identical to the aforementioned in the experiment setting session. From the figure, we observe that FINT displays an superior efficiency by spending the minimum time for each epoch among the ten models, while retaining best prediction performance. The main property of FINT enables the huge speedup: the Hadamard product operations across the features can reduce the problem scale (from exponential to linear).
	
	\subsection{Results from online A/B testing}
	Careful online A/B testing in the advertising display system was conducted. iQiYi, as a large video app, has more than 100 million users using it to watch videos every day. Ads are distributed in multiple locations of the app, including video pre-post ads(video general roll), startup screen ads when the app is opened(open screen), short video feed flow ads (infeed), and long video feed flow ads (semi-feed). During almost a month’s testing, FINT trained with the proposed FINT contributes up to 2.92\% CTR and 3.18\% RPM (Revenue Per Mille) promotion (are shown in Table~\ref{tab.ab}.)
	compared with the baseline models (Wide \& Deep). Now FINT has been deployed online and serves the main traffic.
	\begin{table}[!htbp]
	\centering
		\caption{A/B testing on different advertising positions.}
		\begin{tabular}{ccc}
			\hline
			position      & revenue & click rate \\ \hline
			overall            & +2.72\% & +2.92\%    \\ \hline
			video general roll & +1.53\% & +0.41\%    \\ \hline
			open screen          & +2.67\% & +4.11\%    \\ \hline
			infeed             & +3.39\% & +5.38\%    \\ \hline 
			semi-feed          & +4.81\% & +4.69\%    \\ \hline
		\end{tabular}
		\label{tab.ab}
	\end{table}
	
	\section{Conclusion}
	In this paper, we proposed an efficient and effective CTR predictor named FINT, which aims to learn the high-order feature interactions by employing the Field-aware interaction layer, and captures high-order feature interactions without losing the field-level semantic information. We have conducted extensive experiments on public realistic datasets and A/B testing on large online systems. The obtained results suggest that FINT can learn effective high-order feature interactions, while running faster than state-of-the-art models, meaning a high efficiency on CTR prediction and achieves comparable or even better performances.

\section{Acknowledgement}
This work is partially supported by the major Science and Technology Innovation 2030 ``New Generation Artificial Intelligence'' project 2020AAA0104803.

\bibliographystyle{IEEEbib}
\bibliography{refs}

\end{document}